\documentclass[12pt]{article}

 \usepackage{paralist} 
     \usepackage{placeins} 
     \usepackage{amsmath,amsthm,amssymb}
     \usepackage{epsfig}
     \usepackage{psfrag}
     \usepackage{tabls}
     \usepackage{supertabular}
     \usepackage[square,comma,numbers]{natbib}
     \usepackage[top=2.5cm, bottom=2.5cm, left=2.5cm, right=2.5cm]{geometry}
\usepackage{indentfirst} 
\usepackage{colortbl} 
\definecolor{niceyellow}{rgb}{0.98,0.92,0.73}
\usepackage{fancyhdr}
\usepackage{fancybox}
\usepackage{hyperref} 
    \newcommand{\myfecha}{\today}

\begin{document}
\begin{center}
\Large{\bf  A non-standard approach to introduce simple harmonic motion}
\end{center}
\begin{flushleft}
{\bf Sergio Rojas}\\
srojas@usb.ve \\
{Physics Department, Universidad Sim\'{o}n Bol\'{\i}var,
Ofic. 220, Apdo. 89000, Caracas 1080A, Venezuela.}
\end{flushleft}

{\bf Abstract}.
  We'll be presenting an approach to solve the equation of
simple harmonic motion (SHM) which is non-standard as compared 
with the usual way of solution presented in textbooks. In addition to
help students avoid the unnecessary memorization of formulas
to solve physics problems, this approach could help
instructors to present the subject in a teaching framework
which integrates conceptual and mathematical reasoning,
in a systemic way of thinking that will help students to
reinforce their quantitative reasoning skills by using
mathematical knowledge already familiar to students in a first
calculus-based introductory physics course, such
as the chain rule for derivatives, inverse trigonometric functions,
and integration methods.

\bigskip

\setlength{\parindent}{0.75cm}
\section{\label{sec:1}Introduction\protect}

An interesting account on the teaching and learning of 
math concepts suggest the following:
  ``Teaching advanced algebra in middle school pushes concepts on
students that are beyond normal development at that age.''\cite{Ganem:2009}
Possibly, this observation also applies 
to the teaching of introductory university calculus-based physics courses. 

In fact, consider, for example,
 how the ideas of 
simple harmonic
motion (SHM) are presented in widely used textbooks\cite{TiplerMosca:2003,SerwayJewett:2003,Giancoli:2004,HallidayResnickWalker:2000}.
SHM is introduced in mathematical terms as solving a second
order ordinary non-homogeneous differential
equation, an alien topic to most students attending their first 
introductory calculus-based physics.
To further increase anxiety on students, the standard
procedure to find the solution (which for most students
would be an unimaginable guess that might be
justified afterward if they guessed correctly)
is taught in such a  way that could
cause   
students to
``see mathematical reasoning as a mysterious process that only experts 
with advanced degrees consulting books filled with incomprehensible 
hieroglyphics can fathom.''\cite{Ganem:2009}  

Accordingly, in what follows will be presenting a non-standard 
straightforward approach to introduce SHM as a completely solvable example
of
Newton's second law of motion,
and which 
only requires knowledge 
of the chain rule for derivatives and of
standard integration techniques.
In addition to helping students to fully understand 
the concepts of Newtonian physics,
since along the solution process students also have the opportunity to
practice inverse trigonometric functions, such a straightforward
technique  also offers the advantage of helping them
to find more sense on what they are
 learning in their math courses and how to apply that knowledge.

After consulting several calculus-based introductory physics textbooks (we 
performed
a particular thorough search  of the
well known textbook by Halliday and Resnick, and the historical 
archives of Journals devoted to the teaching of Physics)
we were surprised to find no mention of our alternative approach for
instructors to teach the subject of SHM.

\section{\label{sec:SHM}The Simple Harmonic Motion Equation}

The usual example for presenting SHM is the spring-mass system driven by
a elastic force which is given by Hooke's Law, 
$\vec{\mathbf{F}} = - k \vec{\mathbf{x}}$, where $k$ is the spring
constant and $\vec{\mathbf{x}}$ is the displacement from the spring's
 equilibrium
position.  For vertical oscillations, the equation of motion 
via Newton's second law applied to
a mass $m$ that could oscillate attached at one end of the spring has
the form:
\begin{equation}
m \frac{d^2 z}{dt^2} + k(z-L)=mg,
\label{eq:1}
\end{equation}
where $L$ is the length of the spring, $g$ is the magnitude of 
the local gravitational constant, 
and $z$ is the vertical position, 
with $z=0$ at the top of the
unstretched spring and positive downward. 

To further motivate students, one could mention that equation (1) can also
be found in the study of the small oscillations of a pendulum and in 
the description of the oscillations of partially submerged objects.

\section{\label{sec:SHMsol}Solving the Simple Harmonic Motion equation}

 As mentioned earlier, without much explanation, 
calculus-based introductory physics textbooks introduce the solution
of equation (\ref{eq:1}) as if students were already familiar with 
the subject of non-homogeneous
ordinary differential equations. This is hardly true, and such an
approach causes many troubles not only for students, who might be
 trying
to think hard and quick enough to truly understand
the significance of the fancy term ordinary differential equation,
but also for instructors, who might be having a hard time 
talking about the subject due to timetable constrain.

 Thus, our alternative approach to find the solution of  
equation (\ref{eq:1}) can be started via
the chain rule for derivative by 
considering the one dimensional  speed $v$
as a function of the vertical position
$z$, which allow us to rewrite the second derivative in equation
(\ref{eq:1}) in the form
\begin{equation}
v=\frac{d z}{d t} = v(z(t)) \Longrightarrow 
\frac{d^2 z}{dt^2} = \frac{d}{dt}(\frac{d z}{d t})=\frac{d v}{d t} 
                   = \frac{d v}{d z} \frac{d z}{d t} 
                   = v \frac{d v}{d z}.
\label{eq:2}
\end{equation}
Correspondingly, using this relation in  equation (\ref{eq:1}) one obtains,
\begin{equation}
   v \frac{d v}{d z} = 
     \left(\frac{k}{m}\right)\left( (\frac{m}{k})C - z \right)
\Longrightarrow
   \int_{v=v_0}^{v} v {d v} = 
\int_{z=z_0}^{z} \left(\frac{k}{m}\right)\left( (\frac{m}{k})C - z \right) dz,
\label{eq:3}
\end{equation}
where $C = g + (k/m)L$. Carrying out the integral yields,
\begin{equation}
   v^2 = v_0^2 + 2 C (z-z_0) -  
     \left(\frac{k}{m}\right)\left( z^2 -  z_0^2 \right),
\label{eq:4}
\end{equation}
which can be casted in the forms,
\begin{subequations}
\label{eq:5}
\begin{gather}
   v^2 = v_0^2 + 
     \left(\frac{k}{m}\right)\left( \left(z_0 - \frac{m}{k}C\right)^2
             -  \left(z - \frac{m}{k}C\right)^2 \right)
\label{eq:5a} \\
v^2  + \left(\frac{k}{m}\right) \left(z - \frac{m}{k}C\right)^2  =
 v_0^2 + \left(\frac{k}{m}\right)\left(z_0 - \frac{m}{k}C\right)^2.
\label{eq:5b} 
\end{gather}
\end{subequations}
Arriving at this point, 
one could take advantage of the arrangements of terms  given 
by equation (\ref{eq:5b}) and explain
students about a conserved quantity that holds at every stage of the motion,
which later on will be introduced in the context of energy conservation
for this type of problems.  
In addition,
one could also explore some particular cases for $v_0$ and $z_0$ in order
to help students to get a better grasp of the evolution 
$v$ versus the  position $z$. Moreover, instructors could introduce
the idea of the frequency of the oscillatory motion by considering the
units of $k/m = \omega^2$.

To continue, without loss
of generality equation (\ref{eq:5a}) can be written in the form,
\begin{equation}
   v = \frac{dz}{dt} = 
\sqrt{\frac{k}{m}}\left[(\frac{m}{k})v_0^2 
+ \left(z_0 - \frac{m}{k}C\right)^2
- \left(z - \frac{m}{k}C\right)^2 \right]^{\frac{1}{2}}, 
\label{eq:6}
\end{equation}
from which one has,
\begin{equation}
   \int_{z=0}^{z}\frac{dz}{ 
\left[ b^2  - \left(z - \frac{m}{k}C\right)^2 \right]^{\frac{1}{2}}} 
=
\int_{t=0}^{t}\sqrt{\frac{k}{m}} dt
\label{eq:7}
\end{equation}
on which $b = 
\sqrt{ (\frac{m}{k})v_0^2 + \left(z_0 - \frac{m}{k}C\right)^2}$.
The solution of equation (\ref{eq:7}) has the form,
\begin{equation}
 \arcsin{\left[ \frac{z - \frac{m}{k}C}{b}\right]} 
 - \arcsin{\left[ \frac{z_0 - \frac{m}{k}C}{b}\right]} 
=
\sqrt{\frac{k}{m}} t.
\label{eq:8}
\end{equation}
Now, from this last equation one can obtain,
\begin{equation}
  \frac{z - \frac{m}{k}C}{b}  =
\sin{\left[\sqrt{\frac{k}{m}} t\right]}
\cos{\left[\arcsin{\left[ \frac{z_0 - \frac{m}{k}C}{b}\right]}\right]} + 
\cos{\left[\sqrt{\frac{k}{m}} t\right]}
\sin{\left[\arcsin{\left[ \frac{z_0 - \frac{m}{k}C}{b}\right]}\right]}. 
\label{eq:9}
\end{equation}
To write equation (\ref{eq:9})
 in a familiar form, one needs to use the following relations,
\begin{subequations}
\label{eq:10}
\begin{gather}
  \sin\left[\arcsin{(x)}\right] = x \label{eq:10a}\\
  \cos\left[\arcsin{(x)}\right] = \sqrt{1-x^2}.
\label{eq:10b}
\end{gather}
\end{subequations}
While the identity given by equation (\ref{eq:10a}) could be already familiar 
to students,
equation (\ref{eq:10b}) might not be so. Nevertheless, it can be
introduce by means of the already known trigonometric expression
$\cos{(\theta)} = \sqrt{1-\sin^2{(\theta)}}$. If one lets $x=\sin{(\theta)}$,
then $\theta = \arcsin{(x)}$ and equation (\ref{eq:10b}) follows 
immediately. 
Consequently, using equation (\ref{eq:10}) to rearrange equation (\ref{eq:9})
one obtains the familiar solution,
\begin{equation}
  z = \left( L+\frac{m}{k}g \right) 
+ \left[ v_0 \sqrt{\frac{m}{k}} \right]\sin{\left[\sqrt{\frac{k}{m}} t\right]}
+ 
\left[ z_0 - \left(L+\frac{m}{k}g\right) \right]
\cos{\left[\sqrt{\frac{k}{m}} t\right]},
\label{eq:11}
\end{equation}
where we used that $b^2 - (z_0 - \frac{m}{k}C)^2 = \frac{m}{k} v_0$ and
that $C=g+\frac{k}{m}L$.

Once the solution of equation (\ref{eq:1}) is obtained as 
equation (\ref{eq:11}),
instructors have the freedom to briefly talk about the subject of 
non-homogeneous
ordinary
differential equations, followed by a further exploration of the obtained
solution for the oscillations of a mass attached to a spring from the
physical point of view.   

\section{\label{sec:conclusion}Concluding Remarks}

A non-standard approach  for solving the equation of
motion leading to SHM has been presented. Our interest is to help
students to avoid
solving physics problems merely by applying memorized
formulas, a practice encourage by physics textbook writers  through 
their typical
``formulae summary'' found at the end of each chapter,
a bad habit which unfortunately
can be found even in classroom teaching \cite{Hamed:2008}.

Rather than memorizing an anzat to solve a non-homogeneous ordinary
differential equation, a subject with which generally
students in their  first 
calculus-based introductory physics course are not familiar,  
we make use of the chain rule for derivatives and
of first steps of integration techniques to yield the
solution of the involved equation for SHM. Moreover, while solving 
the equation, 
in addition to practicing the use of inverse
trigonometric functions, students are also introduced to a non-familiar
inverse trigonometry identity. In this way we could encourage them
to be more proactive in applying what they are learning of their 
math courses.

We also note that
research in {\em Physics Education Research}
\cite{Ehrlich:2002,Reif:1981,ReifScott:1999,Yeatts:1992,HeronMeltzer:2005,Heller:1992a}
shows clear evidence that when a problem solving methodology 
\cite{Heller:1992a,Rojas:2010rmf} is applied
via active teaching and learning
strategies \cite{Heller:1992b,Ehrlich:2007,Schoenfeld:1977,Duda:2008},
students' abilities to solve physics problems
quantitatively is strengthened, and their conceptual
understanding of physics is enhanced.
 
To paraphrase Heron and Meltzer,
learning to approach problems in a systematic way starts from teaching
and learning the interrelationships among conceptual knowledge, 
mathematical 
skills and logical reasoning.\cite{HeronMeltzer:2005}
In physics, this necessarily requires the teaching of a good 
deal of mathematical computations.
In such context,
the problem solving methodology proposed here for the SHM 
equation could help
instructors to present the subject in a teaching framework
which integrates conceptual and mathematical reasoning,
which will help
students to
reinforce their quantitative reasoning skills by using
mathematical knowledge already familiar to them,
as could be the chain rule for derivatives, inverse trigonometric functions,
and integration methods.

\bibliographystyle{unsrt}
\bibliography{ajp_ref,cita_bridge,RojasS_submission_ref}
\label{LastPage}
\end{document}